\documentclass[aps,prl,twocolumn,superscriptaddress,showpacs,nofootinbib]{revtex4}
\usepackage[latin1]{inputenc}
\usepackage{amssymb}
\usepackage{amsmath}
\usepackage{array, multirow, hhline, tabularx, graphicx}
\usepackage{bm}
\usepackage{bbm}
\usepackage{textcomp}
\usepackage{ulem}
\usepackage{psfrag}
\usepackage{verbatim}
\usepackage[usenames]{color}
\begin{document}

\title{\bf\large Number-theoretic expressions obtained through analogy between prime factorization and optical interferometry}

\author{Gabriel Seiden}
\affiliation{Department of Environmental Sciences and Energy Research, Weizmann Institute of Science, Rehovot 76100, Israel}
\email{gabriel.seiden@weizmann.ac.il}

\begin{abstract}
Prime factorization is an outstanding problem in arithmetic, with important consequences in a variety of fields, most notably cryptography. Here we employ the intriguing analogy between prime factorization and optical interferometry in order to obtain, for the first time, analytic expressions for closely related functions, including the number of distinct prime factors.
\end{abstract}
\maketitle

For millennia, people have been fascinated with the world of numbers and in particular with prime numbers. As early as $300$ BC Euclid proved the infinitude of the primes and laid the foundation for the proof of the fundamental theorem of arithmetic. The last centuries have seen a surge in scientific activity focused on the primes and related mathematical themes, with leading mathematician such as Euler and Riemann greatly contributing to our current knowledge regarding the complexity of the distribution of primes. 
However, prime factorization of integers, especially that of large numbers, is still a formidable task. As a consequence, different applications in fields such as cryptography have been developed, which take advantage of the great difficulty of factoring large numbers (e.g., the RSA cryptosystem \cite{Rivest1978}).

During the last decades, physicists have contributed to the investigation of integer factorization as well as related mathematical topics, such as the Riemann hypothesis (see \cite{Schumayer2011} and the references therein), through investigating the relationship with various physical systems.
In particular, the intriguing relationship between quadratic Gauss sums and integer factorization has led to recent experimental realization of integer factorization through NMR and optical interference (see e.g., \cite{Vandersypen2001, Mehring2007, Bigourd2008, Tamma2011, Merkel2011}). 

In this Letter we explore the relationship between prime factorization and optical intereferometry with the aim of obtaining novel analytic expressions for number-theoretic functions directly related to prime factorization. Our study, which is based on the relationship between the multiple-slit interference experiment and the occurrence of the primes, exhibits the potential of physical analogy not only in realizing a mathematical enigma experimentally, but also contributing to the theoretical endeavor to unravel the enigma. 

\begin{center}
\begin{table}
\resizebox{6 cm}{5.5 cm}{
\begin{tabular}{|c|c|c|c|c|c|c|c|c|c|c|}
\tableline
n & 2 & 3 & 5 & 7 & 11 & 13 & 17 & 19 & 23 & $\omega(n)$ \\  \tableline
1 & 0 & 0 & 0 & 0 & 0 & 0 & 0 & 0 & 0 & 0\\ \tableline
2 & 1 & 0 & 0 & 0 & 0 & 0 & 0 & 0 & 0 & 1 \\ \tableline
3 & 0 & 1 & 0 & 0 & 0 & 0 & 0 & 0 & 0 & 1 \\ \tableline
4 & 2 & 0 & 0 & 0 & 0 & 0 & 0 & 0 & 0 & 1 \\ \tableline
5 & 0 & 0 & 1 & 0 & 0 & 0 & 0 & 0 & 0 & 1 \\ \tableline
6 & 1 & 1 & 0 & 0 & 0 & 0 & 0 & 0 & 0 & 2 \\ \tableline
7 & 0 & 0 & 0 & 1 & 0 & 0 & 0 & 0 & 0 & 1 \\ \tableline
8 & 3 & 0 & 0 & 0 & 0 & 0 & 0 & 0 & 0 & 1 \\ \tableline
9 & 0 & 2 & 0 & 0 & 0 & 0 & 0 & 0 & 0 & 1 \\ \tableline
10 & 1 & 0 & 1 & 0 & 0 & 0 & 0 & 0 & 0 &  2 \\ \tableline
11 & 0 & 0 & 0 & 0 & 1 & 0 & 0 & 0 & 0 & 1 \\ \tableline
12 & 2 & 1 & 0 & 0 & 0 & 0 & 0 & 0 & 0 & 2 \\ \tableline
13 & 0 & 0 & 0 & 0 & 0 & 1 & 0 & 0 & 0 & 1 \\ \tableline
14 & 1 & 0 & 0 & 1 & 0 & 0 & 0 & 0 & 0 & 2 \\ \tableline
15 & 0 & 1 & 1 & 0 & 0 & 0 & 0 & 0 & 0 & 2 \\ \tableline
16 & 4 & 0 & 0 & 0 & 0 & 0 & 0 & 0 & 0 & 1 \\ \tableline
17 & 0 & 0 & 0 & 0 & 0 & 0 & 1 & 0 & 0 & 1 \\ \tableline
18 & 1 & 2 & 0 & 0 & 0 & 0 & 0 & 0 & 0 & 2 \\ \tableline
19 & 0 & 0 & 0 & 0 & 0 & 0 & 0 & 1 & 0 & 1 \\ \tableline
20 & 2 & 0 & 1 & 0 & 0 & 0 & 0 & 0 & 0 & 2 \\ \tableline
21 & 0 & 1 & 0 & 1 & 0 & 0 & 0 & 0 & 0 & 2 \\ \tableline
22 & 1 & 0 & 0 & 0 & 1 & 0 & 0 & 0 & 0 & 2 \\ \tableline
23 & 0 & 0 & 0 & 0 & 0 & 0 & 0 & 0 & 1 & 1 \\ \tableline
24 & 3 & 1 & 0 & 0 & 0 & 0 & 0 & 0 & 0 & 2 \\ \tableline
25 & 0 & 0 & 2 & 0 & 0 & 0 & 0 & 0 & 0 & 1 \\ \tableline
\end{tabular}}
\caption{Prime factorization of the first twenty-five positive integers. The rightmost column presents the corresponding number of distinct prime factors.}
\label{table:factor}
\end{table}
\end{center}

The fundamental theorem of arithmetic states that any positive integer $n$ can be decomposed into prime numbers in a unique manner, that is:

\begin{equation}
n=\prod_{i=1}^{\omega(n)}{p_i^{\alpha_{p_i}(n)}},
\end{equation}
where $\omega(n)$ is the number of distinct primes function, $p_i$ is the the $i^{th}$ prime and $\alpha_{p_i}(n)$ is the corresponding power. 
Table \ref{table:factor} depicts the integer decomposition into primes of the first twenty-five integers. From the table one can deduce three features of prime factorization. First, one notes that each prime appears periodically as $n$ increases, with a period equal to the corresponding prime. This is a consequence of the fact that for a given prime $p$ all integers divisible by $p$ can be written as $N_p=mp$, where m is an integer.  We define the following function, which conveys the periodicity of the occurrence of a prime $p$:

\begin{equation}
\label{eq:alpha}
\psi_p(n)=
\begin{cases}
1~~~~\alpha_p(n)>0\\
0~~~~otherwise\\
\end{cases}.
\end{equation}
The second feature concerns with the {\it synchronization} in the occurrence of the different primes in Table \ref{table:factor}, or equivalently, the synchronization of the $\psi_p(n)$'s. A close examination shows that the occurrence is {\it in phase} with respect to extrapolating the periodic set $\{\psi_p(n)\}$ backward toward zero, where $\psi_p(0)=1$ for all primes. The third feature we note is that a particular prime number $p^*$ appears in the table when $\psi_p(0)=0$ for all $p<p^*$. Thus, knowledge of an initial set of $m$ primes ${2,3,...,p_m}$ allows for the determination of at least  one consecutive prime. In fact, it is easy to be convinced that this set allows for the determination of all consecutive primes obeying: $p\le p_m^2$.

\begin{figure}[ht]
\includegraphics[width=8.5cm]{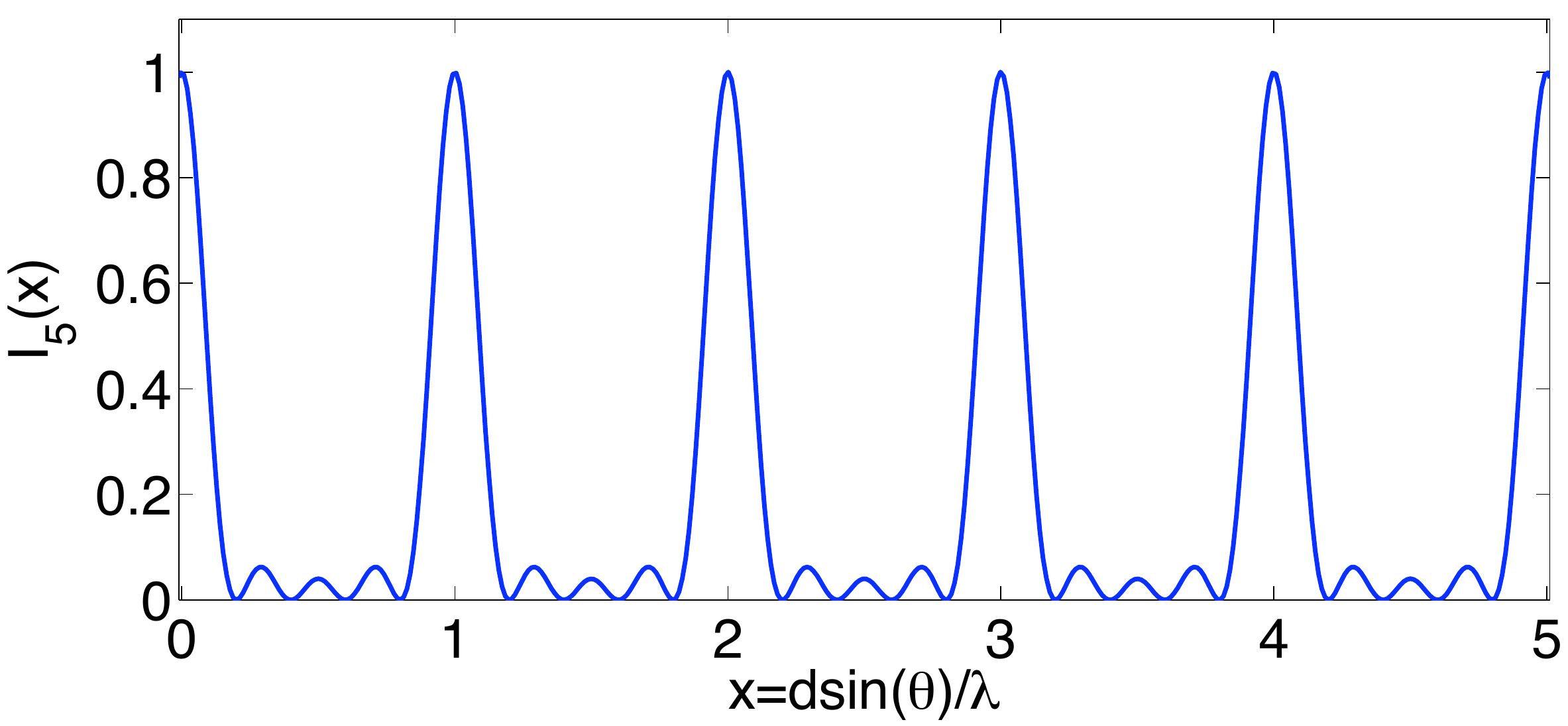}
\caption{Normalized intensity corersponding to a set of five monochromatic, coherent narrow slits.}
\label{fig:I_5}
\end{figure}

Examining Eq. \ref{eq:alpha}, we recall a similar functional dependency which appears in optical interferometry. It is well known that the normalized intensity of $p$ coherent, monochromatic sources in the form of infinitely narrow slits positioned on a line with spacing $d$ is given by \cite{Lipson1995},

\begin{equation}
\label{eq:Ip}
I_p(\theta)=\frac{1}{p^2}\frac{sin^2\left(p\pi d \sin(\theta)/\lambda\right)}{sin^2\left(\pi d \sin(\theta)/\lambda\right)}.
\end{equation}
where, $\lambda$ and $\theta$ are the wavelength and angle with the normal at the midpoint of the slit arrangement, respectively. Figure \ref{fig:I_5} depicts the intensity for the particular case $p=5$, where we have used: $x=dsin(\theta)/\lambda$. The function $I_p(x)$ can be made to coincide for all positive integer values with ${\psi_p(n)}$ through the coordinate transformation $x'=xp$. This is analogous to scaling the distance between sources by $p$ in Eq. \ref{eq:Ip}, a scaling which will prove valuable further on when we superimpose contributions of different (non-coherent) sets of sources. Thus, $I_p(x')$ has its maximal value of unity for any integer divisible by $p$, and vanishes for all other integers. 
In what follows we will omit the superscript '. 

\begin{figure}[ht]
\includegraphics[width=8.5cm]{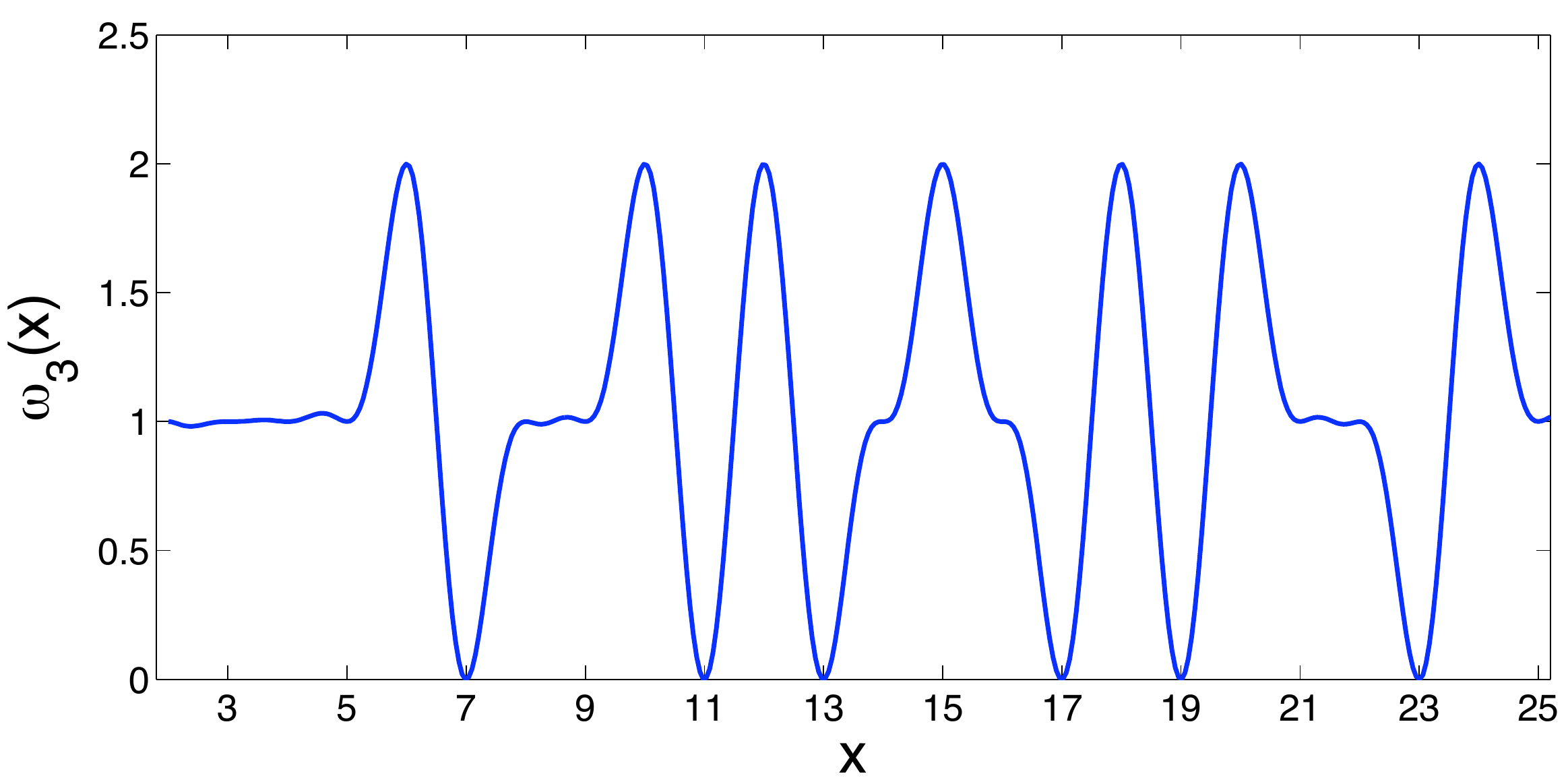}
\caption{Functional dependence of $\omega_3(x)$ ($p_3=5$), which represents the intensity corresponding to the set of sources depicted in Fig. \ref{fig:sources_rev}. The zeros in the interval $5< x\le 25$ are located at the consecutive primes: 7, 11, 13 17, 19 and 23.}
\label{fig:omega_3}
\end{figure}

The fact that $\psi_p(n)=I_p(n)$ enables us to make the following statement with respect to the third feature pointed out earlier. Let ${2,3,...,p_m}$ be a set of $m$ initial primes and define:

\begin{equation}
\label{eq:omega_m}
\omega_m(x)=\sum_{i=1}^m{I_{p_i}(x)}=\sum_{i=1}^m{\frac{1}{p_i^2}\frac{sin^2\left(\pi x\right)}{sin^2\left(\pi x/p_i\right)}}.
 \end{equation}
Then finding the consecutive set of primes which obey  $p<p_m^2$ is equivalent to finding the zeros of $\omega_m(x)$ in the interval $p_m< x\le p_m^2$. This is a direct consequence of the fact $I_p(x)$ obtains its zeros when the numerator vanishes but the denominator does not. Thus, $I_p(x)=0$ for any integer not divisible by $p$, and therefore $\omega_m(x)$ obtains its zeros for any integer which is not divisible by the first $m$ primes. Figure \ref{fig:omega_3} depicts the case $m=3$ ($p_3=5$), for which the zeros of $\omega_3(x)$ in the interval $5< x\le25$ are located at $7,11,13,17,19$ and $23$. 

\begin{figure}[ht]
\includegraphics[width=7cm]{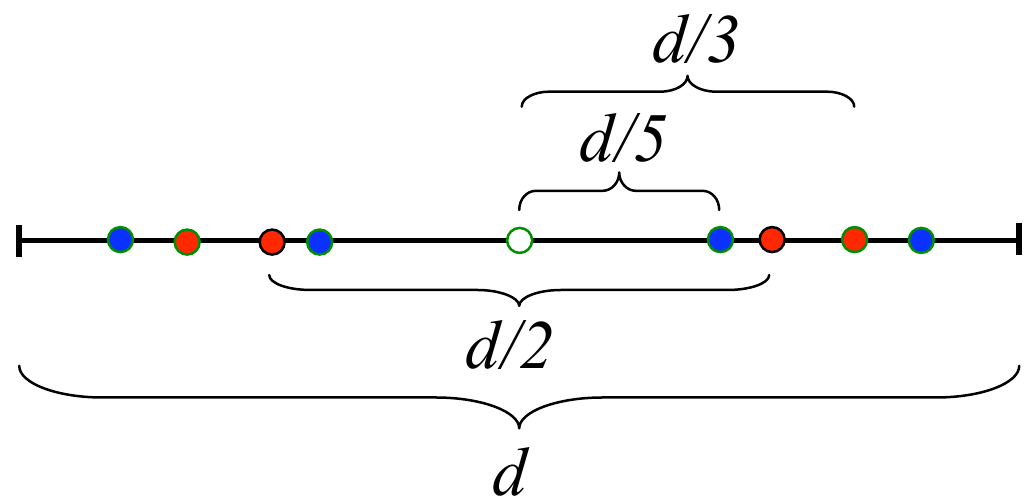}
\caption{Arrangement of narrow slit sources (represented by circles) on a line segment of length $d$. The sources are divided into three sets of $2$ (red), $3$ (green) and $5$ (blue) sources with spacings $d/2$, $d/3$ and $d/5$, respectively. The only overlapping point is the midpoint, shared by all odd sets (white circle).}
\label{fig:sources_rev}
\end{figure}

The analogy with interference of narrow slit sources is conceptually rather simple, though experimental implementation might prove difficult. Equation \ref{eq:omega_m} represents a superposition of contributions due to $m$ sets of infinitely narrow slit sources (in an experimental realization one would need to take into consideration the effect of diffraction due to the finite width of the slits). Each such set has $p_i$ coherent sources with spacing $d/p_i$. While all sources (in all sets) have the same wavelength $\lambda$, different sets are non-coherent. 
Figure \ref{fig:sources_rev} depicts the arrangement of narrow slit sources corresponding to the $m=3$ scenario.  Note that due to the fact that the superposition is of sets consisting of a prime number of sources, the sources never overlap, except for the middle point (marked white in Fig. \ref{fig:sources_rev}), which is shared by all odd sets. 
We also note that if the different sets (i.e., pertaining to different prime numbers) were coherent, which would be equivalent to illuminating the slit arrangement with a single monochromatic, coherent beam, the functional dependency of the intensity would be qualitatively different, in large part due to the coupling terms representing interference between the different sets.

The partial sum $\omega_m(x)$ has three qualitatively different regions. For $2\le x\le p_m$,  $\omega_m(n)$ faithfully represents the number of distinct primes in the factorization of $n$. In particular, $\omega_m(n)=1$ for $n=p_i$. In the second region, $p_m<x\le p_m^2$ the zeros of $\omega_m(x)$ coincide with the consecutive set of primes ($p_m<p\le p_m^2$). In the third region $p_m^2<x$, $\omega_m(n)$ in general neither represents the number of distinct prime factors, nor has its zeros at locations of prime numbers. 

There is an intriguing connection between $\omega_m(n)$ and the sieve of Erastosthenes, which provides a rigorous procedure for finding all primes smaller or equal to a given integer $N$. Starting from the first prime $p_1=2$, one marks all multiples of the prime up to $N$. The first integer left unmarked at every step is by definition a prime number and is used in the next step in order to locate additional composite numbers. 
The procedure terminates when one has located all primes smaller or equal to $\sqrt{N}$.  
In a similar manner, as pointed out above, the partial sum $\omega_m(n)$ distinguishes between composite integers ($\omega_m(n)>0$) and primes ($\omega_m(n)=0$) in the interval $p_m<n\le p_m^2$, based on a known initial set of $m$ primes. 
Thus, by locating the zeros of $\omega_m(n)$ in this interval one obtains the consecutive primes $p_{m+1}, ..., p_{m+k}$ (where $p_{m+k}$ is the closest prime smaller than $p_m^2$). The next step in the procedure is to locate the zeros of $\omega_{m+k}(n)$ in the interval $(p_{m+k},p_{m+k}^2$). As in the sieve of Erastosthenes, the procedure ends when one has located the largest prime, which is smaller or equal to $\sqrt{N}$. 

An important consequence of Eq. \ref{eq:omega_m} is:
\begin{equation}
\omega(n)=\lim_{m\rightarrow\infty}\omega_m(n)=\sum_{i=1}^{\infty}{\frac{1}{p_i^2}\frac{sin^2\left(\pi n\right)}{sin^2\left(\pi n/p_i\right)}}.
\label{fig:omega_limit}
\end{equation}
It is important to note that, apart from being important in its own right (e.g., \cite{Hardy1917, Erdos1940}), $\omega(n)$ is related to other important functions encountered in number theory (and elsewhere), such as the prime-counting function $\pi(x)$ and Riemann's zeta function $\zeta(s) ~(s>1)$. In particular, one has the following identity \cite{Hardy1979},   

\begin{equation*}
\sum_{n=1}^{\infty}{\frac{2^{\omega(n)}}{n^s}}=\frac{\zeta^2(s)}{\zeta(2s)}.
\label{eq:zeta}
\end{equation*}

The analogy between prime factorization and optical interferometry can be taken even further. 
An important function related to prime factorization is the total number of prime factors (with multiplicity) defined as,
\begin{equation}
\Omega(n)=\sum_{i=1}^{\omega(n)}{\alpha_{p_i}(n)}.
\label{eq:Omega}
\end{equation}
In order to obtain an expression for $\Omega(n)$, we are led to examine the dependence of $\alpha_p$ over the positive integers which are divisible by $p$ (for all other integers $\alpha_p=0$). We have already seen that $I_p(n)$ conveys the periodicity of occurrence of the prime $p$ in the factorization of integers. 

\begin{figure}[ht]
\includegraphics[width=8.5cm]{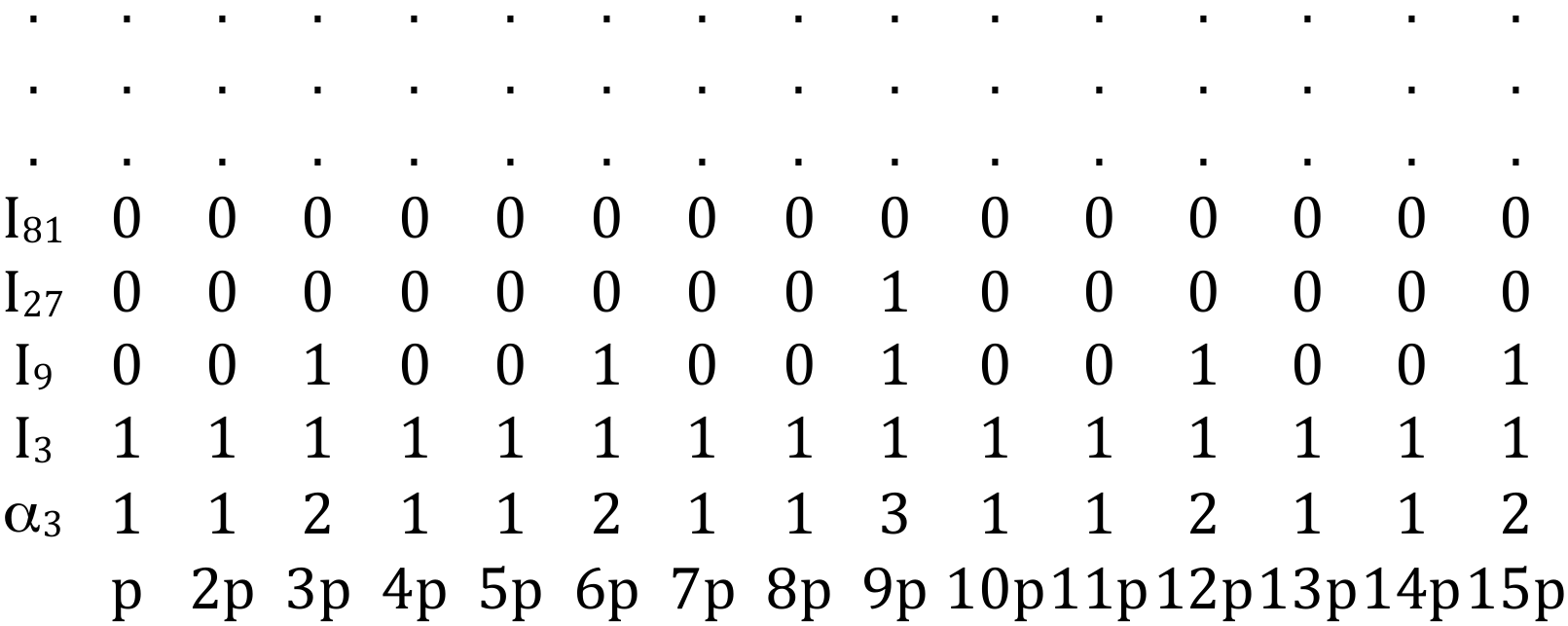}
\caption{Relationship between $\alpha_p(n)$ and the set $\{I_{p^m}(n) \}$ (m=1,2,3,...) for the particular case $p=3$.}
\label{fig:alpha_3}
\end{figure}

Figure \ref{fig:alpha_3} depicts the first fifteen non-vanishing entries for the particular case $\alpha_3(n)$, together with the corresponding values of $I_3,~I_{9}$, $I_{27}$ and $I_{81}$. A close examination leads to the conclusion that $\alpha_p$  is a superposition of contributions from: $I_p,~I_{p^2},~I_{p^3}, ...$, 

\begin{equation}
\alpha_p(n)=\sum_{j=1}^{\infty}{I_{p^j}(n)}=\sum_{j=1}^{\infty}{\frac{1}{p^{2j}}\frac{sin^2\left(\pi n\right)}{sin^2\left(\pi n/p^j\right)}}.
\label{eq:alpha_2}
\end{equation}
Substituting Eq. \ref{eq:alpha_2} into Eq. \ref{eq:Omega} one then gets:
\begin{equation}
\Omega(n)=\sum_{i=1}^{\omega(n)}{\sum_{j=1}^{\infty}{I_{p_i^j}(n)}}=\sum_{i=1}^{\omega(n)}{\sum_{j=1}^{\infty}{\frac{1}{p_i^{2j}}\frac{sin^2\left(\pi n\right)}{sin^2\left(\pi n/p_i^j\right)}}}.
\end{equation}

In summary, we have explored the analogy between prime factorization of integers and optical interferometry with the aim of obtaining analytic expressions for number-theoretic functions closely related to prime factorization. It was shown that the normalized intensity corresponding to a set of $p$ coherent sources possesses the functional dependency pertaining to the occurrence of the corresponding prime in the factorization of the integers. We then examined the partial sum $\omega_m(n)$, pertaining to the intensity field of $m$ sets, each consisting of a prime number of narrow slits, and elaborated on its connection with the sieve of Erastosthenes. An analytic expression for the number of distinct prime factors $\omega(n)$ was then derived through taking the limit of infinite sets of slits. Finally, we also obtained analytic expressions for the functions $\alpha_p(n)$ and $\Omega(n)$, which together with $\omega(n)$ completely define prime factorization.


\begin{thebibliography}{10}%
\makeatletter
\providecommand \@ifxundefined [1]{%
 \ifx #1\undefined \expandafter \@firstoftwo
 \else \expandafter \@secondoftwo
\fi
}%
\providecommand \@ifnum [1]{%
 \ifnum #1\expandafter \@firstoftwo
 \else \expandafter \@secondoftwo
\fi
}%
\providecommand \enquote [1]{``#1''}%
\providecommand \bibnamefont  [1]{#1}%
\providecommand \bibfnamefont [1]{#1}%
\providecommand \citenamefont [1]{#1}%
\providecommand\href[0]{\@sanitize\@href}%
\providecommand\@href[1]{\endgroup\@@startlink{#1}\endgroup\@@href}%
\providecommand\@@href[1]{#1\@@endlink}%
\providecommand \@sanitize [0]{\begingroup\catcode`\&12\catcode`\#12\relax}%
\@ifxundefined \pdfoutput {\@firstoftwo}{%
 \@ifnum{\z@=\pdfoutput}{\@firstoftwo}{\@secondoftwo}%
}{%
 \providecommand\@@startlink[1]{\leavevmode}%
 \providecommand\@@endlink[0]{}%
}{%
 \providecommand\@@startlink[1]{%
  \leavevmode
  \pdfstartlink
   attr{/Border[0 0 1 ]/H/I/C[0 1 1]}%
   user{/Subtype/Link/A<</Type/Action/S/URI/URI(#1)>>}%
  \relax
 }%
 \providecommand\@@endlink[0]{\pdfendlink}%
}%
\providecommand \url  [0]{\begingroup\@sanitize \@url }%
\providecommand \@url [1]{\endgroup\@href {#1}{\urlprefix}}%
\providecommand \urlprefix [0]{URL }%
\providecommand \Eprint[0]{\href }%
\@ifxundefined \urlstyle {%
  \providecommand \doi [1]{doi:\discretionary{}{}{}#1}%
}{%
  \providecommand \doi [0]{doi:\discretionary{}{}{}\begingroup
  \urlstyle{rm}\Url }%
}%
\providecommand \doibase [0]{http://dx.doi.org/}%
\providecommand \Doi[1]{\href{\doibase#1}}%
\providecommand \bibAnnote [3]{%
  \BibitemShut{#1}%
  \begin{quotation}\noindent
    \textsc{Key:}\ #2\\\textsc{Annotation:}\ #3%
  \end{quotation}%
}%
\providecommand \bibAnnoteFile [2]{%
  \IfFileExists{#2}{\bibAnnote {#1} {#2} {\input{#2}}}{}%
}%
\providecommand \typeout [0]{\immediate \write \m@ne }%
\providecommand \selectlanguage [0]{\@gobble}%
\providecommand \bibinfo [0]{\@secondoftwo}%
\providecommand \bibfield [0]{\@secondoftwo}%
\providecommand \translation [1]{[#1]}%
\providecommand \BibitemOpen[0]{}%
\providecommand \bibitemStop [0]{}%
\providecommand \bibitemNoStop [0]{.\EOS\space}%
\providecommand \EOS [0]{\spacefactor3000\relax}%
\providecommand \BibitemShut [1]{\csname bibitem#1\endcsname}%
\bibitem{Rivest1978}%
  \BibitemOpen
  \bibfield{author}{%
  \bibinfo {author} {\bibfnamefont{R.~L.}~\bibnamefont{Rivest}}, \bibinfo {author}
  {\bibfnamefont{A.}~\bibnamefont{Shamir}},\ and\ \bibinfo {author}
  {\bibfnamefont{L.}~\bibnamefont{Adleman}},\ }%
  \bibfield{journal}{%
  \bibinfo {journal} {Comm. ACM}\ }%
  \textbf{\bibinfo {volume} {21}},\ \bibinfo {pages} {120} (\bibinfo {year}
  {1978})%
  \bibAnnoteFile{NoStop}{Rivest1978}%
\bibitem{Schumayer2011}%
  \BibitemOpen
  \bibfield{author}{%
  \bibinfo {author} {\bibfnamefont{D.}~\bibnamefont{Schumayer}}\ and\ \bibinfo
  {author} {\bibfnamefont{D.~A.~W.}\ \bibnamefont{Hutchinson}},\ }%
  \bibfield{journal}{%
  \bibinfo {journal} {Rev. Mod. Phys.}\ }%
  \textbf{\bibinfo {volume} {83}},\ \bibinfo {pages} {307} (\bibinfo {year}
  {2011})%
  \bibAnnoteFile{NoStop}{Schumayer2011}%
\bibitem{Vandersypen2001}%
  \BibitemOpen
  \bibfield{author}{%
  \bibinfo {author} {\bibfnamefont{L.~M.~K.}\ \bibnamefont{Vandersypen}}, \bibinfo
  {author} {\bibfnamefont{M.}~\bibnamefont{Steffen}}, \bibinfo {author}
  {\bibfnamefont{G.}~\bibnamefont{Breyta}}, \bibinfo {author}
  {\bibfnamefont{C.~S.}\ \bibnamefont{Yannoni}}, \bibinfo {author}
  {\bibfnamefont{M.~H.}\ \bibnamefont{Sherwood}},\ and\ \bibinfo {author}
  {\bibfnamefont{I.~L.}\ \bibnamefont{Chuang}},\ }%
  \bibfield{journal}{%
  \bibinfo {journal} {Nature}\ }%
  \textbf{\bibinfo {volume} {414}},\ \bibinfo {pages} {883} (\bibinfo {year}
  {2001})%
  \bibAnnoteFile{NoStop}{Vandersypen2001}%
\bibitem{Mehring2007}%
  \BibitemOpen
  \bibfield{author}{%
  \bibinfo {author} {\bibfnamefont{M.}~\bibnamefont{Mehring}}, \bibinfo
  {author} {\bibfnamefont{K.}~\bibnamefont{M\"uller}}, \bibinfo {author}
  {\bibfnamefont{I.~Sh.}\ \bibnamefont{Averbukh}}, \bibinfo {author}
  {\bibfnamefont{W.}~\bibnamefont{Merkel}},\ and\ \bibinfo {author}
  {\bibfnamefont{W.~P.}\ \bibnamefont{Schleich}},\ }%
  \bibfield{journal}{%
  \bibinfo {journal} {Phys. Rev. Lett.}\ }%
  \textbf{\bibinfo {volume} {98}},\ \bibinfo {pages} {120502} (\bibinfo {year}
  {2007})%
  \bibAnnoteFile{NoStop}{Mehring2007}%
\bibitem{Bigourd2008}%
  \BibitemOpen
  \bibfield{author}{%
  \bibinfo {author} {\bibfnamefont{D.}~\bibnamefont{Bigourd}}, \bibinfo
  {author} {\bibfnamefont{B.}~\bibnamefont{Chatel}}, \bibinfo {author}
  {\bibfnamefont{W.~P.}\ \bibnamefont{Schleich}},\ and\ \bibinfo {author}
  {\bibfnamefont{B.}~\bibnamefont{Girard}},\ }%
  \bibfield{journal}{%
  \bibinfo {journal} {Phys. Rev. Lett.}\ }%
  \textbf{\bibinfo {volume} {100}},\ \bibinfo {pages} {030202} (\bibinfo {year}
  {2008})%
  \bibAnnoteFile{NoStop}{Bigourd2008}%
\bibitem{Tamma2011}%
  \BibitemOpen
  \bibfield{author}{%
  \bibinfo {author} {\bibfnamefont{V.}~\bibnamefont{Tamma}}, \bibinfo {author}
  {\bibfnamefont{H.}~\bibnamefont{Zhang}}, \bibinfo {author}
  {\bibfnamefont{X.}~\bibnamefont{He}}, \bibinfo {author}
  {\bibfnamefont{A.}~\bibnamefont{Garuccio}}, \bibinfo {author}
  {\bibfnamefont{W.~P.}\ \bibnamefont{Schleich}},\ and\ \bibinfo {author}
  {\bibfnamefont{Y.}~\bibnamefont{Shih}},\ }%
  \bibfield{journal}{%
  \bibinfo {journal} {Phys. Rev. A}\ }%
  \textbf{\bibinfo {volume} {83}},\ \bibinfo {pages} {020304(R)} (\bibinfo {year}
  {2011})%
  \bibAnnoteFile{NoStop}{Tamma2011}%
\bibitem{Merkel2011}%
  \BibitemOpen
  \bibfield{author}{%
  \bibinfo {author} {\bibfnamefont{W.}~\bibnamefont{Merkel}}, \bibinfo {author}
  {\bibfnamefont{S.}~\bibnamefont{W\"olk}}, \bibinfo {author}
  {\bibfnamefont{W.~P.}\ \bibnamefont{Schleich}}, \bibinfo {author}
  {\bibfnamefont{I.~Sh.}\ \bibnamefont{Averbukh}}, \bibinfo {author}
  {\bibfnamefont{B.}~\bibnamefont{Girard}},\ and\ \bibinfo {author}
  {\bibfnamefont{G.~G.}\ \bibnamefont{Paulus}},\ }%
  \bibfield{journal}{%
  \bibinfo {journal} {New J. Phys.}\ }%
  \textbf{\bibinfo {volume} {13}},\ \bibinfo {pages} {103008} (\bibinfo {year}
  {2011})%
  \bibAnnoteFile{NoStop}{Merkel2011}%
\bibitem{Lipson1995}%
  \BibitemOpen
  \bibfield{author}{%
  \bibinfo {author} {\bibfnamefont{S.~G.}\ \bibnamefont{Lipson}}, \bibinfo
  {author} {\bibfnamefont{H.}~\bibnamefont{Lipson}},\ and\ \bibinfo {author}
  {\bibfnamefont{D.~S.}\ \bibnamefont{Tannhauser}},\ }%
  {\bibinfo {title} {Optical Physics}}\ (\bibinfo {publisher} {Cambridge
  University Press},\ \bibinfo {year} {1995})%
  \bibAnnoteFile{NoStop}{Lipson1995}%
\bibitem{Hardy1917}%
  \BibitemOpen
  \bibfield{author}{%
  \bibinfo {author} {\bibfnamefont{G.~H.}\ \bibnamefont{Hardy}}\ and\ \bibinfo
  {author} {\bibfnamefont{S.}~\bibnamefont{Ramanujan}},\ }%
  \bibfield{journal}{%
  \bibinfo {journal} {Quart. J. Math.}\ }%
  \textbf{\bibinfo {volume} {48}},\ \bibinfo {pages} {76} (\bibinfo {year}
  {1917})%
  \bibAnnoteFile{NoStop}{Hardy1917}%
\bibitem{Erdos1940}%
  \BibitemOpen
  \bibfield{author}{%
  \bibinfo {author} {\bibfnamefont{P.}~\bibnamefont{Erd\"os}}\ and\ \bibinfo
  {author} {\bibfnamefont{M.}~\bibnamefont{Kac}},\ }%
  \bibfield{journal}{%
  \bibinfo {journal} {Am. J. Math.}\ }%
  \textbf{\bibinfo {volume} {62}},\ \bibinfo {pages} {738} (\bibinfo {year}
  {1940})%
  \bibAnnoteFile{NoStop}{Erdos1940}%
\bibitem{Hardy1979}%
  \BibitemOpen
  \bibfield{author}{%
  \bibinfo {author} {\bibfnamefont{G.~H.}\ \bibnamefont{Hardy}}\ and\ \bibinfo
  {author} {\bibfnamefont{E.~M.}\ \bibnamefont{Wright}},\ }%
  {\bibinfo {title} {An Introduction to the theory of numbers}}\ (\bibinfo
  {publisher} {Oxford University Press},\ \bibinfo {year} {1979})%
  \bibAnnoteFile{NoStop}{Hardy1979}%
\end{thebibliography}
\end{document}